# Recombination Amplitude Calculations of Noble Gases, in Length and Acceleration Forms, beyond Strong Field Approximation.

Siddharth Bhardwaj[1*], Sang-Kil Son[2], Kyung-Han Hong[1], Chien-Jen Lai[1], Franz X Kärtner[1,2,3] and Robin Santra[2,3]

[1]*Department of Electrical Engineering and Computer Science and Research Laboratory of Electronics, Massachusetts Institute of Technology, 77 Massachusetts Ave, Cambridge MA 02139*
[2]*Center for Free-Electron Laser Science, DESY, Notkestraße 85, D-22607 Hamburg, Germany*
[3]*Dept. of Physics, University of Hamburg, 20355 Hamburg, Germany*

**Abstract:** *Transition of an electron from a free to a bound state is critical in determining the qualitative shape of the spectrum in high-order harmonic generation (HHG), and in tomographic imaging of orbitals. We calculate and compare the recombination amplitude, from a continuum state described by a plane wave and an outgoing scattering eigenstate, to the bound state for the noble gases that are commonly used in HHG. These calculations are based on the single active electron model and the Hartree-Fock-Slater method, using both the length form and the acceleration form of the dipole matrix element. We confirm that the recombination amplitude versus emitted photon energy strongly depends upon the wavefunction used to describe the free electron. Depending on the choice of the wavefunction and the dipole form, the square of the absolute value of the recombination amplitude can differ by almost two orders of magnitude near the experimentally measured Cooper minima. Moreover, only the outgoing scattering eigenstates with the length form roughly predict the experimentally observed Cooper minimum for Ar (~50 eV) and Kr (~85 eV). We provide a detailed derivation of the photorecombination cross sections (PRCSs) from photo ionization cross sections (PICSs) calculated by the relativistic random phase approximation (RRPA)., For Ar, Kr and Xe, compare the total PICSs calculated using our recombination amplitudes with that obtained from RRPA. We find that PICS calculated using the outgoing scattering eigenstates with the length form is in better agreement with the RRPA calculations than the acceleration form.*

## I. Introduction

The three step model (TSM) is commonly used to describe the dynamics of an electron in the strong field regime, which is responsible for generation of high-order harmonics [1] . In this semi-classical description, the dynamics of a single electron is simplified into three distinct steps: ionization, propagation and recombination (back to the orbital of origin in the parent atom or molecule). The amplitude of the harmonic dipole is determined by a product of the amplitudes of each of the three steps [2]. It has been shown that the qualitative shape of the plateau in the HHG spectrum almost exclusively depends on the recombination amplitude [3] [4] [5]**.** Additionally, the recombination step serves as a probe that imprints information about electronic orbital [6]**,** atomic attosecond dynamics [7] and molecular motion [8] onto the harmonic spectrum. The central role that the recombination step plays in the aforementioned experiments serves as a strong motivation for a systematic study of the recombination amplitude of noble gases commonly used in HHG.

The recombination amplitude describes the transition of the returning electron back into the atomic orbital from where it originated. The strength of this transition is given by a dipole transition matrix element that depends on the wavefunction used to describe the returning electron. Since, it is not possible to calculate the exact many-body eigenstates of the electrons in the atom, one needs to resort to various approximations to describe the electronic wavefunction. A key assumption made in the TSM is that only a single electron participates in the HHG process while the ion core remains frozen. In this picture, the electron from the outermost valence orbital aligned along the laser polarization tunnel ionizes and upon return recombines to the same orbital. This is called the single active electron approximation (SAEA). In this paper, we use an effective atomic potential ($V_{HFS}$) to calculate the bound and continuum eigenstates. The effective atomic potential is obtained from the Hartree Fock Slater (HFS) model which employs a local density approximation for the exact exchange interaction [9]. An important consequence of approximating the exact many-body eigenstate with the eigenstates of $V_{HFS}$ while using the exact hamiltonian in the acceleration form, is that the form invariance of the dipole operator is lost, i.e., the recombination amplitude depends whether the dipole operator is in the length, or in the acceleration form [10].

Another important assumption made in the TSM is that after ionization the electron moves only under the influence of the laser field without any interaction with the Coulomb potential of the ion core. The rationale behind this assumption, often referred to as the strong field approximation (SFA), is that in strong-field processes like HHG, the ionized electron can travel hundreds of Bohr radii away from the atom. Therefore, its trajectory, for the most part, is that of a free electron in an external electric field which can be described by Volkov states (plane waves with time-dependent momentum) [11]. The basic assumptions of SFA are: (a) neglect the laser field for the calculation of bound states and (b) neglect the core Coulomb potential for the calculation of the continuum states [12]. Recently measured HHG spectrum of Ar is shown to have a deep minimum (related to the Cooper minimum of its photoelectron spectrum) that is independent of the laser intensity or wavelength [13] [14]. This Cooper minimum of the HHG spectrum can be theoretically reproduced if the ionized electron is defined by outgoing scattering eigenstates [10] rather than plane waves while keeping the bound states as eigenstates of field-free Coulomb potential [13]. This indicates that while the first assumption of SFA appears to be valid, the second assumption is not accurate. Hence we are motivated to use the outgoing scattering eigenstates rather than the plane waves in the calculation of the recombination amplitudes of all noble gases used in HHG.

Since photorecombination and photoionization cross sections have the same dipole transition matrix element, the recombination amplitude can be compared with the extensively studied photoionization cross section (PICS). The mathematical relation between the recombination amplitude and the PICS will be discussed in detail in Section II. The central potential model with a single active electron has also been used to calculate the PICSs in the extreme ultraviolet regime ($0 \sim 100\ eV$) [15] and x-ray regime [16] which are in qualitative agreement with experimental results. However, this simple model does not take into account inter-channel coupling needed to explain the PICS of 3$p$ shell in Ar and 4$d$ shell in Xe [17]. Techniques such as R-matrix theory [18], random phase approximation with exchange (RPAE) [19] incorporate inter-channel coupling as a perturbation, while the relativistic random phase approximation (RRPA) [20] [21], in addition, also includes the relativistic effects. PICSs calculated using the RRPA match very well with the experimental measurements [22]. As we will see in Section II, it is possible to calculate the photorecombination cross section (PRCS) from PICS. In principal, by comparing the *PRCS* obtained from RRPA with the PRCS obtained from our recombination amplitude calculation, one can discuss the limitations of the central potential model with a single active electron. However, due to lack of $m_J$ resolved PICS data from RRPA, we compare the total photoionization cross section from our theory and from RRPA.

In this paper, we extensively investigate the recombination amplitudes of the commonly used noble gases in HHG. We show that the recombination amplitude versus emitted photon energy critically depends upon the choice of the wavefunction used to describe the returning electron as well as the form of the dipole operator. In some cases, the square of the absolute value of the recombination amplitude can differ by two orders of magnitude because the Cooper minima are located at different energies. This is critical when predicting the efficiency of HHG process and in attosecond pulse generation at certain photon energies. In order to show the limitation of the central potential model, we compare the PICSs calculated using our recombination amplitudes with PICS obtained from the RRPA. This paper is structured as follows: in Section 2 we derive the recombination amplitude and show how to calculate PRCS from PICS. In Section 3 the results are discussed and compared with the PICS data calculated using RRPA [23].

## II. Theory

The recombination amplitude or the dipole matrix element of transition from the momentum normalized free state $|\Psi_k\rangle$ to the bound state $|\Psi_g\rangle$ can be written in the length and the acceleration form:

$$a_{len}(k) = \langle \Psi_g | z | \Psi_k \rangle, \tag{1a}$$

$$a_{acn}(k) = \langle \Psi_g | -\partial_z V | \Psi_k \rangle. \tag{1b}$$

When used in the TSM, Eq. (1a) gives the dipole moment [1] and Eq. (1b) gives the dipole acceleration [24]. The recombination amplitude in the length and the acceleration form are related by:

$$\langle \Psi_g | -\partial V_z | \Psi_k \rangle = \omega_{gk}^2 \langle \Psi_g | z | \Psi_k \rangle, \tag{2}$$

where

$$\omega_{gk} = \frac{k^2}{2} + I_p, \tag{3}$$

is the energy of the photon emitted after recombination, $k$ is the momentum of the ejected electron and $I_p$ is the ionization potential. Although we use an approximate Hartree-Fock-Slater potential $V_{HFS}$ to calculate the electronic eigenstates, for the calculation of the recombination amplitude in the acceleration form, the exact multi-electron potential $V$ is used (See Eq. (9) of [25]). Since the electron-electron interaction term cancels out, we get

$$\left\langle \Psi_g \left| -\partial V_z \right| \Psi_k \right\rangle = -Z_N \left\langle \Psi_g \left| \frac{z}{r^3} \right| \Psi_k \right\rangle \tag{4}$$

where $Z_N$ is the atomic number. Validity of Eq. (2) is predicated upon the usage of exact many-electron wavefunction for bound and free states.

We begin the calculation of the recombination amplitude by expanding the plane wave in the spherical co-ordinate system as an infinite sum of free spherical waves. For simplicity, it is assumed that the ionized electron moves along the z-direction. This allows us to limit the expansion to spherical waves with $m_l = 0$. Then the momentum-normalized plane wave and the ground state, projected on the **r** space become :

$$\left\langle \mathbf{r} \middle| \Psi_k^{pl} \right\rangle = \sum_{l=0}^{\infty} a_l \frac{u_{kl}^{pl}(r)}{r} Y_{l0}(\Omega_r), \quad a_l = \frac{i^l}{2k}\sqrt{\frac{2l+1}{\pi}} \tag{5a}$$

$$\left\langle \mathbf{r} \middle| \Psi_g \right\rangle = \frac{u_g(r)}{r} Y_{l0}(\Omega_r) \tag{5b}$$

The radial part $u_{kl}^{pl}(r)$ is proportional to the momentum-normalized spherical Bessel function of the first kind:

$$u_{kl}^{pl} = \sqrt{\frac{2}{\pi}} kr \, j_l(kr) \tag{6}$$

$Y_{l0}(\Omega_r)$ is the spherical harmonic with zero magnetic quantum number. The radial part of the ground state orbital is calculated by solving the Hartree-Fock-Slater eigenvalue problem using a generalized pseudospectral method [26] [27] on a non-uniform grid. Recombination amplitudes for the plane waves are calculated by inserting Eq. (5a) and (5b) in Eq. (1a) for the length form and Eq. (1b) for the acceleration form. The calculation is simplified because the summation over all angular momenta is reduced to the terms that satisfy the dipole selection rule $\Delta l = \pm 1$. Then, the recombination amplitude of a plane wave into the outermost orbital of He ($|l = 0, m_l = 0\rangle$) in the length and acceleration form are

$$a_{len}^{pl}(k) = a_1 c_1 \left\langle u_g \middle| r \middle| u_{k1}^{pl} \right\rangle \tag{7a}$$

$$a_{acn}^{pl}(k) = -a_1 c_1 Z_N \left\langle u_g \middle| \frac{1}{r^2} \middle| u_{k1}^{pl} \right\rangle. \tag{7b}$$

For other noble gases where the outermost orbital is $|l = 1, m_l = 0\rangle$, the recombination amplitude of the plane wave in the length and the acceleration form are:

$$a_{len}^{pl}(k) = a_0 c_0 \langle u_g | r | u_{k0}^{pl} \rangle + a_2 c_2 \langle u_g | r | u_{k2}^{pl} \rangle \tag{7c}$$

$$a_{acn}^{pl}(k) = -a_0 c_0 Z_N \langle u_g | \frac{1}{r^2} | u_{k0}^{pl} \rangle - a_2 c_2 Z_N \langle u_g | \frac{1}{r^2} | u_{k2}^{pl} \rangle. \tag{7d}$$

Here, $a_l$ is the coefficient of expansion in Eq. (5a), $c_l = \langle Y_{lg}^{m=0} | \cos\theta | Y_l^{m=0} \rangle$ is the angular part of the integral and $u_g(r)$ is the radial part of the ground state orbital. Using $\cos\theta = \sqrt{\frac{4\pi}{3}} Y_{10}$, we can express $c_l$ in terms of Wigner 3j symbol [28]:

$$\int Y_{l_1 m_1}(\theta,\phi) Y_{l_2 m_2}(\theta,\phi) Y_{l_3 m_3}(\theta,\phi) \sin\theta \, d\theta d\phi = \sqrt{\frac{(2l_1+1)(2l_2+1)(2l_2+1)}{4\pi}} \begin{pmatrix} l_1 & l_2 & l_3 \\ 0 & 0 & 0 \end{pmatrix} \begin{pmatrix} l_1 & l_2 & l_3 \\ m_1 & m_2 & m_3 \end{pmatrix} \tag{8}$$

For a spherically symmetric potential, the outgoing scattering eigenstate can be obtained by replacing the radial part of the free spherical wave $u_{kl}^{pl}(r)$ in Eq. (7) by the radial part of the corresponding partial wave $e^{i(\delta_l+\sigma_l)} u_{kl}^{sc}(r)$ [29]. The radial part of the continuum states is solved by the fourth-order Runge-Kutta method on a uniform grid [15][30] using the Hartree-Fock-Slater potential determined for the ground state. Similar to the case of the plane wave, the recombination amplitude in the length and the acceleration form of the outgoing scattering eigenstate for He (Eq. (9a) and (9b)) and other noble gases (Eq. (9c) and (9d)) are

$$a_{len}^{sc}(k) = a_1 c_1 e^{i(\delta_1+\sigma_1)} \langle u_g | r | u_{k1}^{sc} \rangle \tag{9a}$$

$$a_{acn}^{sc}(k) = -a_1 c_1 Z_N e^{i(\delta_1+\sigma_1)} \langle u_g | \frac{1}{r^2} | u_{k1}^{sc} \rangle \tag{9b}$$

$$a_{len}^{sc}(k) = a_0 c_0 e^{i(\delta_0+\sigma_0)} \langle u_g | r | u_{k0}^{sc} \rangle + a_2 c_2 e^{i(\delta_2+\sigma_2)} \langle u_g | r | u_{k2}^{sc} \rangle \tag{9c}$$

$$a_{acn}^{sc}(k) = -a_0 c_0 Z_N e^{i(\delta_0+\sigma_0)} \langle u_g | \frac{1}{r^2} | u_{k0}^{sc} \rangle - a_2 c_2 Z_N e^{i(\delta_2+\sigma_2)} \langle u_g | \frac{1}{r^2} | u_{k2}^{sc} \rangle. \tag{9d}$$

In the asymptotic limit, the radial part of the partial wave and the free spherical wave become:

$$u_{kl}^{sc}(r) \underset{r\to\infty}{\to} \sqrt{\frac{2}{\pi}} \sin\left(kr - \frac{l\pi}{2} - \eta \ln 2kr + \sigma_l + \delta_l\right) \tag{10}$$

$$u_{kl}^{pl}(r) \underset{r\to\infty}{\to} \sqrt{\frac{2}{\pi}} \sin\left(kr - \frac{l\pi}{2}\right) \tag{11}$$

In the asymptotic limit, the radial part of the partial wave (Eq. (10)) and the radial part of the free spherical wave (Eq. (11)) differ by a phase shift which is composed of three terms: the $r$ dependent phase term is due to the long-range nature of the Coulomb potential, $\sigma_l$ is the Coulomb phase shift and $\delta_l$ is the phase shift against the regular coulomb wave (due to the short-range part of the HFS potential $V_{HFS}$) [31]. The two terms in Eq. (9c) and (9d) correspond to $s$ and $d$ partial waves that satisfy the dipole transition rule for the bound $p$ orbital. Interplay between the two terms determines the minima in the recombination amplitude, an example of which is the location of the commonly observed Cooper minimum in Ar [13] [14] [32]. Once we have the recombination amplitude, the PRCS for recombining into the orbital $|l=1, m_l=0\rangle$ ($|l=0, m_l=0\rangle$ for He) can be calculated by [28]

$$\frac{d\sigma^R_{len}}{d\Omega_k d\Omega_n} = \frac{4\pi^2\omega^3}{c^3 k}\left|a^{sc}_{len}(k)\right|^2 \tag{12a}$$

$$\frac{d\sigma^R_{acn}}{d\Omega_k d\Omega_n} = \frac{4\pi^2\omega}{c^3 k}\left|a^{sc}_{acn}(k)\right|^2. \tag{12b}$$

Here, $c$ is the speed of light, $k$ is the momentum of the returning electron, and $\omega$ is the angular frequency of the released photon. $\Omega_n$ and $\Omega_k$ are solid angles in the direction of polarization of emitted photon and electron momentum respectively. We want to compare our PRCS with that available in literature. Since PICS have been extensively studied, both experimentally and theoretically, we will first review the method of converting PICS into PRCS.

The reverse of the recombination step described in Section II is the process where a photon with polarization along a solid angle $\Omega_n$ ionizes an electron in the polarization direction from the outermost orbital ($|n, l=0, m_l=0\rangle$ for He, $|n, l=1, m_l=0\rangle$ for other noble gases). Due to this symmetry, the cross sections of the two processes are related by principle of detailed balancing [33]:

$$\frac{d^2\sigma^R}{\omega^2 d\Omega_n d\Omega_k} = \frac{d^2\sigma^I}{k^2 c^2 d\Omega_k d\Omega_n} \tag{13}$$

where $\omega$ is the photon frequency, and $k$ is the electron momentum and $\sigma^R$ is the photorecombination cross section and $\sigma^I$ is the photoionization cross sections respectively. In order to apply the above relation to calculate the PRCS from the total PICS we need (a) the partial PICS which is the contribution of the polarization-aligned outermost orbital, and (b) the differential PICS which is the photoionization cross section of emitting an electron in a given solid angle. The differential PICS can be calculated using [4]

$$\frac{d\sigma^I_{lm_l}}{d\Omega_k} = \frac{\sigma^I_{lm_l}}{4\pi}\left(1+\beta_{lm_l}(\varepsilon)P_2\left(\cos(\theta_k)\right)\right), \tag{14}$$

where $\sigma^I_{lm_l}$ is the partial PICS of $|lm_l\rangle$ orbital, $\beta_{lm_l}$ is the energy dependent asymmetry parameter and $P_2(\cos\theta_k)$ is the second-order Legendre polynomial. The polar angle $\theta_k$ is the angle between the laser polarization and the direction of the ejected electron. Due to the dipole approximation Eq. (14) is independent of the azimuthal angle $\varphi$. Since we are interested in ionization along the laser polarization, $\theta_k$ is set to zero.

### III. Results and Discussion

In Fig. (1), the square of the absolute values of the recombination amplitude - calculated using plane waves and outgoing scattering eigenstates and dipole moment in both the length form and the acceleration form - have been plotted for Ar and Kr. In order to compare the length and the acceleration forms, the former has been multiplied by a pre-factor as shown in Eq. (2). Calculations using the plane waves and the outgoing scattering eigenstates differ by almost two orders of magnitude around the experimentally measured Cooper minima (~50 eV for Ar and ~85 eV for Kr) because the location of the minima predicted by the plane wave is way off from the experimentally measured values. The results are also dependent on the form of the dipole operator: for outgoing scattering eigenstates, the minima for the length and the acceleration forms are located at 44 eV and 86 eV respectively for Argon, and 68 eV and 235 eV respectively for Krypton. The plane wave fails to reproduce the experimentally observed minima irrespective of the form of the dipole operator. This suggests that using a plane wave to describe the returning electron is a poor approximation which has also been demonstrated in the calculation of HHG spectrum from molecules using quantum rescattering theory [28]. Hence, for the rest of the paper, we will only focus on outgoing scattering eigenstates.

For outgoing scattering eigenstates, we have compared the square of the absolute values of the recombination amplitude of various noble gases. In Fig. (2) and Fig. (3) these comparisons are shown for the length form and for the acceleration form respectively. The plots of different gases vary significantly as a function of the emitted photon energy. This information is crucial in determining the choice of gas for HHG in a particular energy range. The effect of the choice of the dipole form on the recombination amplitude of a given gas can be observed by comparing the plots in Fig. (2) and Fig. (3). For He and Ne, the results are quantitatively similar. As we move to heavier gases, the effect of the choice of the dipole form on the recombination amplitude becomes evident due to form-dependent minima. Therefore, we need to determine which form of the dipole moment is more suitable in the modeling of HHG.

In modeling of HHG using the TSM, the acceleration form is often preferred because in the calculation of macroscopic propagation of HHG, the dipole acceleration is proportional to the polarization term in Maxwell's Equation. Usage of the length form would require taking a double time derivative which can become numerically cumbersome in 3-D modeling of the HHG process . Two reasons have been put forth in favor of the acceleration form. First, it has been shown that the high harmonic spectrum of hydrogen obtained from TSM is in better agreement with exact time dependent Schrödinger equation when the recombination amplitude in the acceleration form is used [24]. Second, experimentally observed scaling of HHG intensity with the atomic number of the noble gas has been explained using acceleration form in its exact form as shown in Eq. (4) [25]. Due to the presence of atomic number $Z_N$ in Eq. (4), heavier atoms will have a higher recombination amplitude and therefore a stronger HHG radiation.

In both of the aforementioned studies (ref [24] and [25]), preference for the acceleration form stems from the fact that the returning electron is described by a plane wave rather than an outgoing scattering eigenstate. For hydrogen, when an outgoing scattering eigenstate is used to describe the returning electron, the recombination amplitude is form invariant. Similarly, PICS calculated using outgoing scattering eigenstate (of the effective central potential) and length form, increases for heavier gases [15]. Since PRCS is proportional to PICS (Eq. (13)), the former should also increase for heavier gases which explains the increase in HHG yield with atomic number. Hence, it is unclear if the acceleration form is inherently better than the length form. As discussed in the introduction, the lack of form invariance is due to the limitations of the SAE model based on a central potential. Since this approximation is extensively used to model HHG, it is important to know which of the two dipole forms can better reproduce the experimental results and in which energy regimes. In order to do so we will compare the total PICS obtained from our HFS model with the total PICS obtained using RRPA.

Experimentally, the total PICS of noble gases has been extensively studied [34] [35]. The differential and the partial PICS has also been measured for 2s and 2p orbitals of Ne [36], and 3s and 3p orbitals of Ar [37]. Theoretical calculation of the partial PICS and the asymmetry parameter has been done using non-relativistic random phase approximation (RPA) and RRPA [23]. While RPA is quite successful in the calculation of total PICS, it fails to accurately calculate the partial cross section and the asymmetry parameter where the relativistic effects are important [21]. RRPA, on the other hand, includes correlation effects and relativistic spin-orbit coupling and reproduces the experimentally measured asymmetry parameters [38]. Moreover, it also exhibits form invariance [23] [35]. Therefore, we compare the total photoionization cross section from our theory with that obtained from RRPA [23].

In [23], partial PICS ($\sigma_J^I$) of Ar, Kr and Xe have been calculated, where $J$ is the total angular momentum is a constant of motion in the presence of spin-orbit coupling. Partial PICS calculated using SAEA ($\sigma_{l,m_l}^I$) are in the $|l, m_l\rangle$ basis where $l$ is the orbital angular momentum and $m_l$ its component along momentum direction. For our purposes, we need to compare, the total PICS in $\sigma_{l,m_l}^I$ basis $\left[ 2(\sigma_{l=1,m_l=-1}^I + \sigma_{l=1,m_l=0}^I + \sigma_{l=1,m_l=1}^I) \right]$ with total PICS in $\sigma_J^I$ basis $\left[ \sigma_{J=\frac{3}{2}}^I + \sigma_{J=\frac{1}{2}}^I \right]$.

Total PICSs calculated from our model have been compared to the total PICSs obtained from RRPA in Fig. (4), Fig. (5) and Fig. (6) for Ar, Kr and Xe respectively. PICS calculated using the length form and the RRPA results agree fairly well for Ar and Kr. In the case of Xe, the RRPA predicts the experimentally observed "giant resonance" [4]. Our model, based on SAE and dipole moment in the length and acceleration forms, cannot capture this effects because it does not take into account the inter-channel coupling where the conventional TSM breaks down. This tells us that the TSM with a single active electron, which has served so well in predicting the qualitative shape of the HHG spectrum, cannot be used in high-harmonic spectroscopy when multi-electron effects become important.

**IV. Conclusion**

We have calculated the recombination amplitude, in the length and the acceleration forms of the dipole operator, for both plane waves and outgoing scattering eigenstates of atomic HFS potential. We have shown that plane wave approximation fails to predict the Cooper minimum of Ar and Kr. However, these features can be reproduced when the outgoing scattering eigenstates with the dipole moment in the length form are used. We have also shown that the dipole moment in the length form is better than the acceleration form in the calculation of recombination amplitude for certain energies depending upon the noble gas. The comparison with the PICS obtained from existing RRPA calculations reveals that the SAE model has its limitations and more sophisticated theoretical tools are needed to explain the HHG spectrum over all energy range.

**Acknowledgement**

We thank Dr. Stefan Pabst for helpful discussion. This work is supported by FA9550-09-1-0212, FA9550-10-0063 and FA9550-12-1-0080, and Center for Free-Electron Laser Science, DESY.

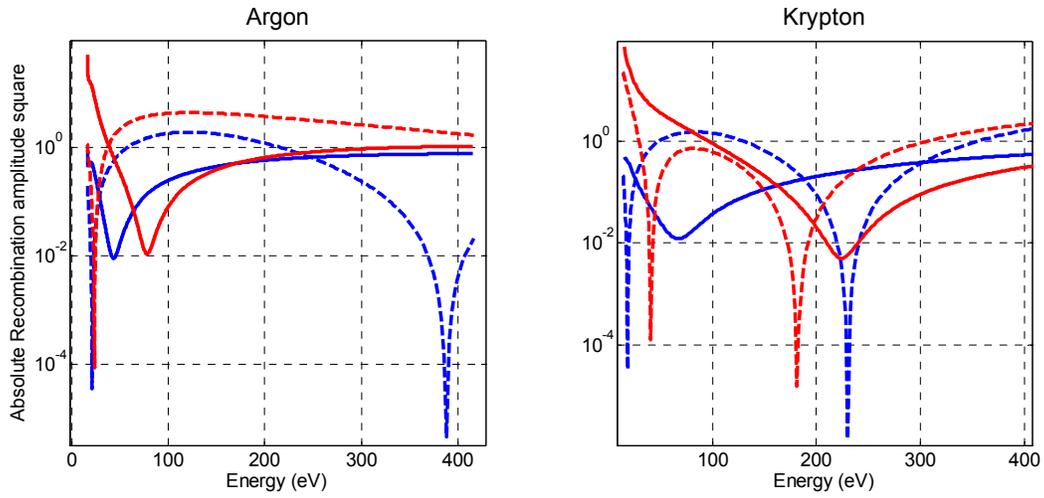

**Figure 1**: Square of absolute values of the Recombination Amplitude of Argon and Krypton for Plane Wave (PW) and Scattering Eigenstate (SC) in Length Form (LF) and Acceleration Form (LF): blue dashed (PW-LF), red dashed (PW-AF), blue solid (SC-LF) and red solid (SC-AF). In order to compare the length and the acceleration form, the former has been multiplied by square of the pre-factor in Eq. (2).

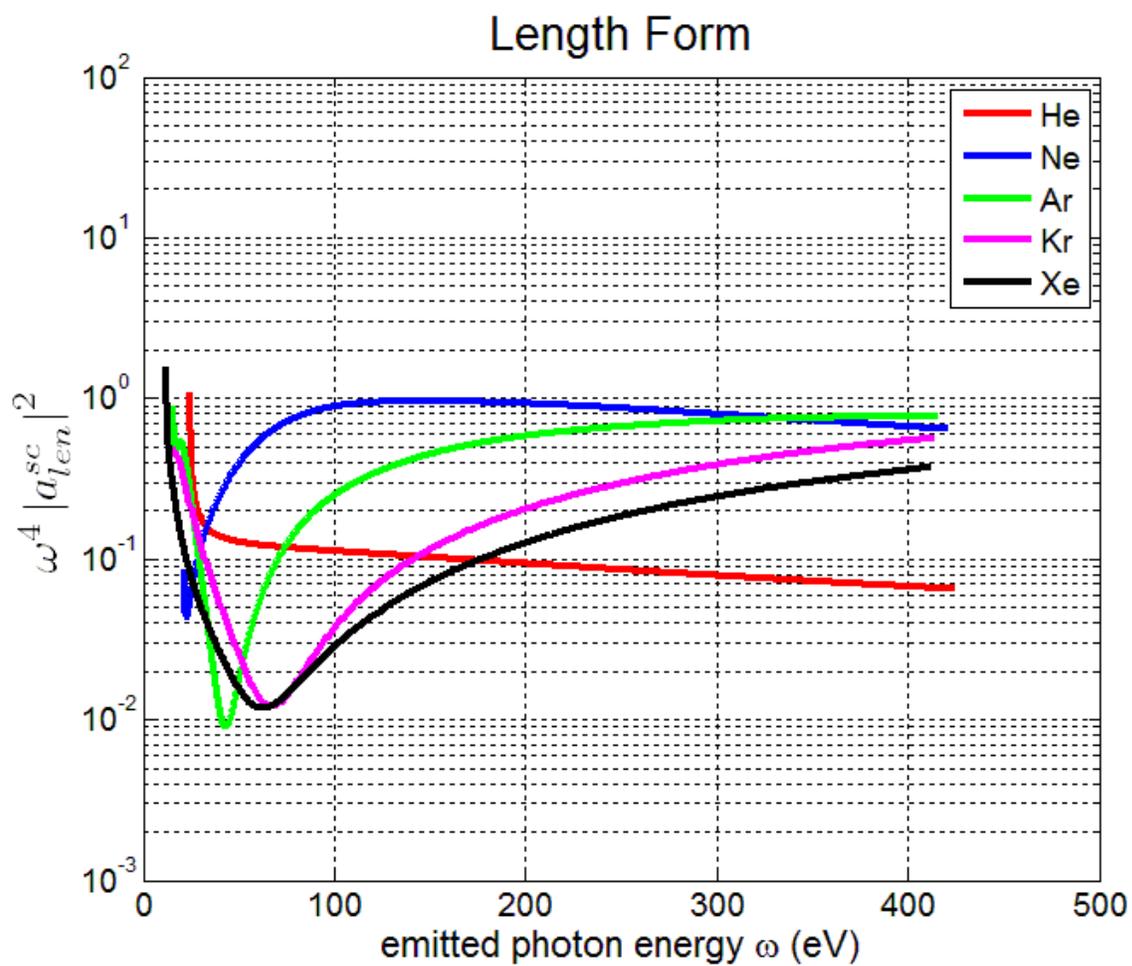

**Figure 2:** Square of the absolute value of the recombination amplitude for outgoing scattering eigenstates in the length form.

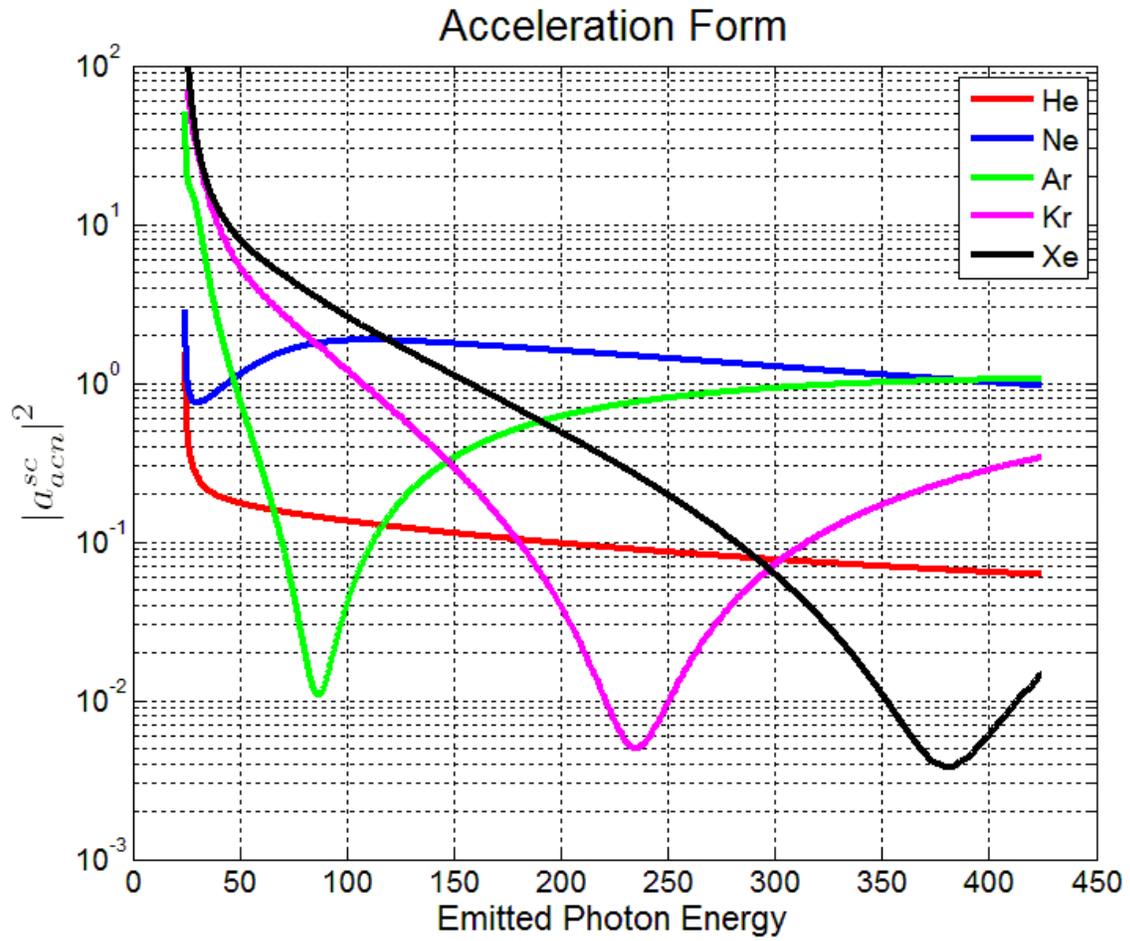

**Figure 3:** Square of the absolute value of the recombination amplitude for outgoing scattering eigenstate in the acceleration form.

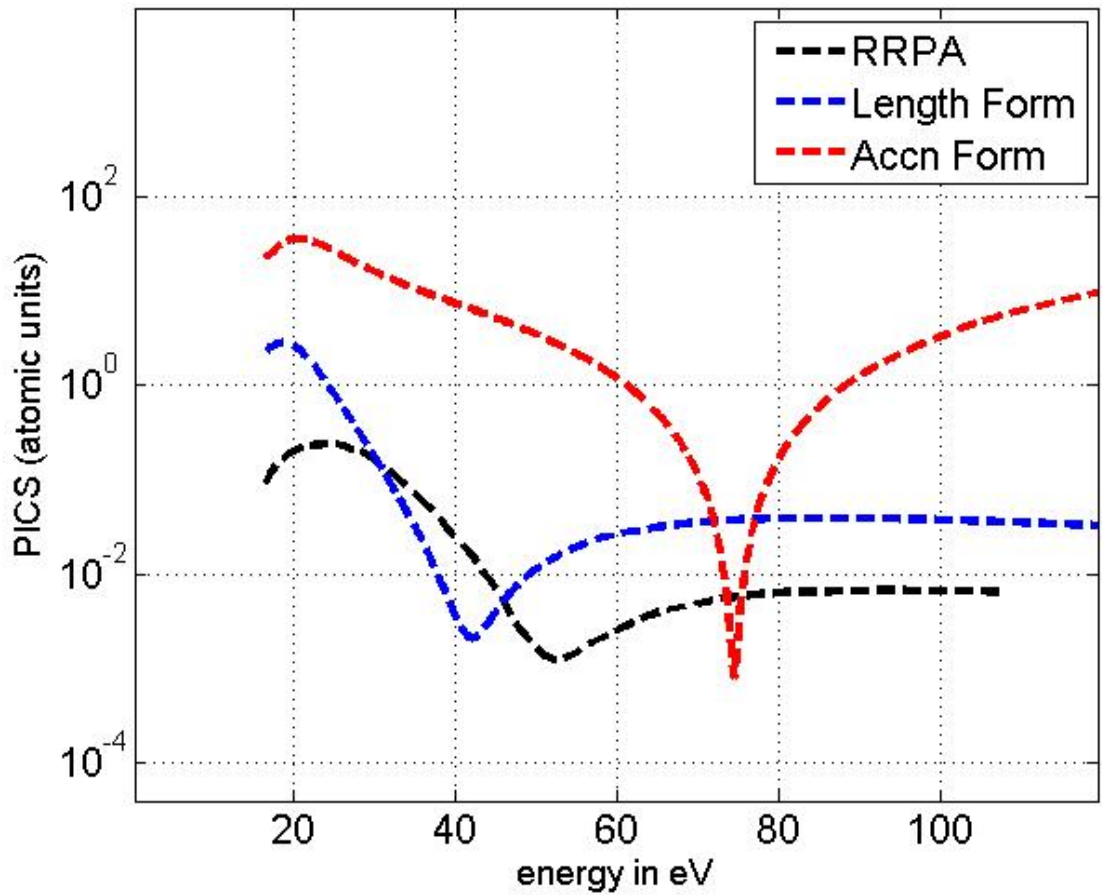

**Figure 4:** Argon's total Photoionization cross sections (PICSs) calculated using outgoing scattering eigenstates with dipole moment in the length form (blue) and the acceleration form (red); and RRPA (black). PICS obtained from the length form is in better agreement with the RRPA calculation ($1\ au^2 = 28Mb$).

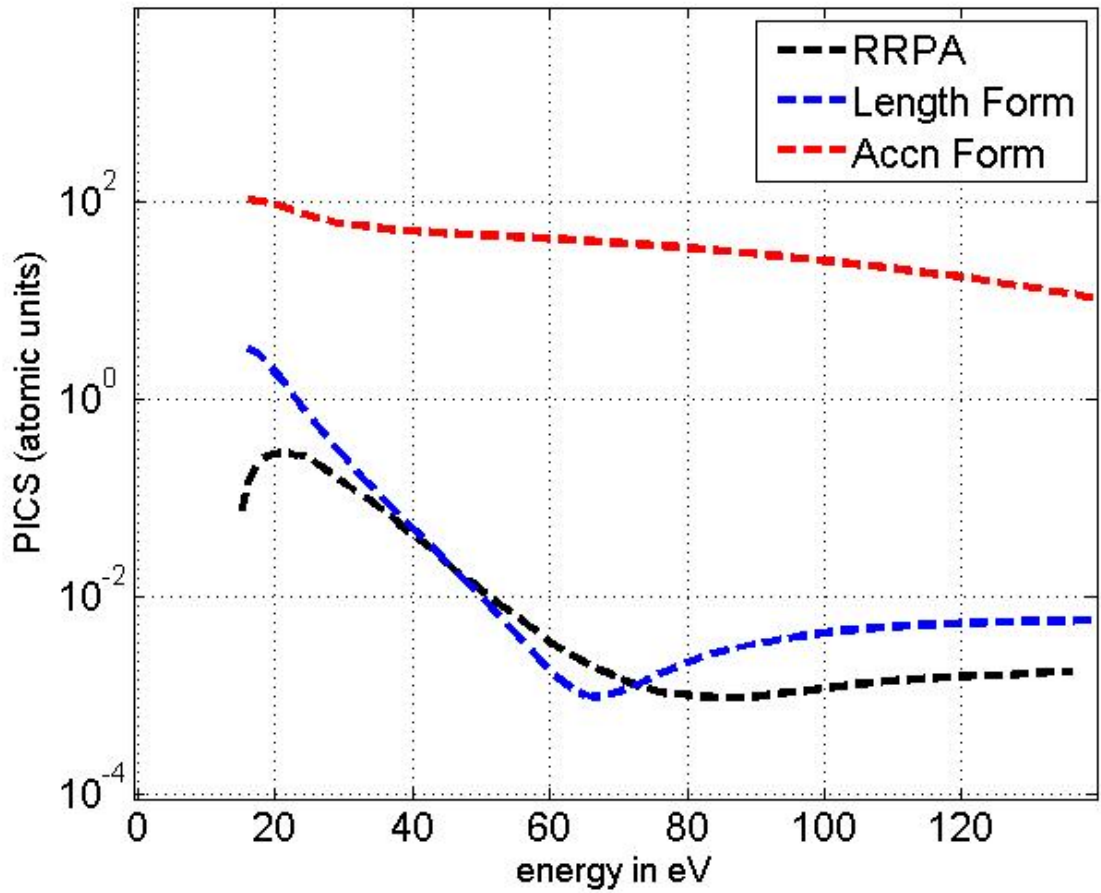

**Figure 5:** Krypton's total Photoionization cross sections (PICSs) calculated using outgoing scattering eigenstates with dipole moment in the length form (blue) and the accelaration form (red); and RRPA (black). PICS obtained from the length form is in good agreement with RRPA calculation in 30 eV to 80 eV range. In the same range, the acceleration form is off by about 4 orders of magnitude ($\mathbf{1\ au^2 = 28Mb}$).
15

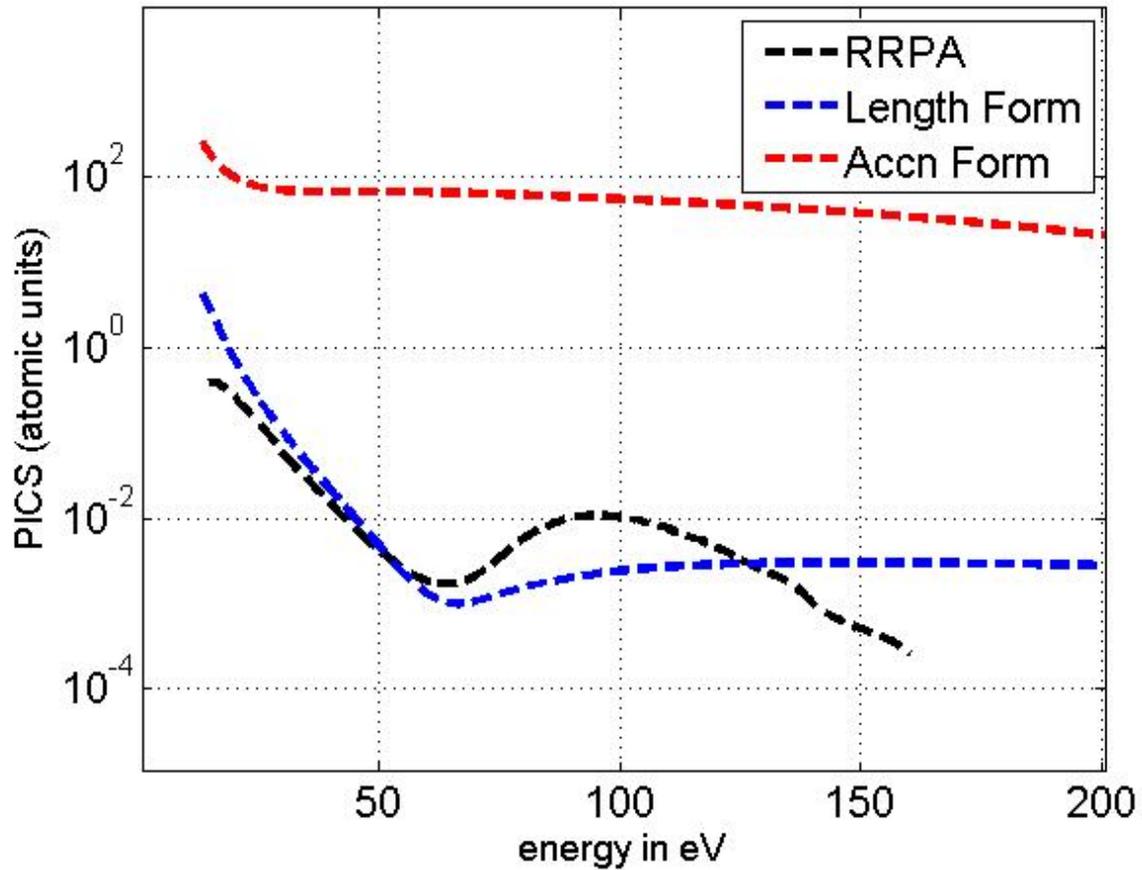

**Figure 6:** Xenon's total Photoionization cross sections (PICSs) calculated using outgoing scattering eigenstates with dipole moment in the length form (blue) and the accelaration form (red); and RRPA (black). PICS obtained from the dipole form is in good agreement with RRPA calculation. In the same range, the acceleration form is off by about 4 orders of magnitude ($\mathbf{1\ au^2 = 28Mb}$).